\documentclass[twocolumn,pre,showpacs]{revtex4}

\usepackage{amsfonts}
\usepackage{amssymb}
\usepackage{amsmath}
\usepackage{graphicx}

\input{tcilatex}

\begin{document}

\title{Langevin approach to synchronization of hyperchaotic time-delay
dynamics}
\author{Adri\'{a}n A. Budini}
\affiliation{Consejo Nacional de Investigaciones Cient\'{\i}ficas y T\'{e}cnicas, Centro
At\'{o}mico Bariloche, Av. E. Bustillo Km 9.5, (8400) Bariloche, Argentina,
and Consortium of the Americas for Interdisciplinary Science and Department
of Physics and Astronomy, University of New Mexico, Albuquerque, New Mexico
87131, USA }
\date{\today }

\begin{abstract}
In this paper, we characterize the synchronization phenomenon of
hyperchaotic scalar non-linear delay dynamics in a fully-developed chaos
regime. Our results rely on the observation that, in that regime, the
stationary statistical properties of a class of hyperchaotic attractors can
be reproduced with a linear Langevin equation, defined by replacing the
non-linear delay force by a delta-correlated noise. Therefore, the
synchronization phenomenon can be analytically characterized by a set of
coupled Langevin equations. We apply this formalism to study anticipated
synchronization dynamics subject to external noise fluctuations as well as
for characterizing the effects of parameter mismatch in a hyperchaotic
communication scheme. The same procedure is applied to second order
differential delay equations associated to synchronization in
electro-optical devices. In all cases, the departure with respect to perfect
synchronization is measured through a similarity function. Numerical
simulations in discrete maps associated to the hyperchaotic dynamics support
the formalism.
\end{abstract}

\maketitle

\section{Introduction}

In the last decades, \textit{synchronization} of chaotic dynamics became a
subject that has attracted of lot of attention \cite%
{book,fujisaka,pikovsky1,pikovsky2,grassberger,pecora,carroll}. In fact,
from a theoretical point of view, this phenomenon seems to contradict the
inheriting sensibility to initial conditions of chaotic dynamics. On the
other hand, the interest in this kind of phenomenon comes from the
possibility of using the unpredictable chaotic trajectories as a carrier
signal in communication channels. In this context, high dimensional systems
with multiple positive Lyapunov exponents, i.e., \textit{hyperchaotic
dynamics} \cite{rossler}, have been proposed as a resource for improving the
security in the communication schemes. Synchronization of hyperchaotic
systems has therefore also become an area of active research \cite%
{kocarev,peng,laiR,ali,parker}.

Chaotic dynamics described by differential delay equations arise in the
description of many different kind of situations, such as physiology \cite%
{mackey,foss}, biology \cite{donald}, economy \cite{economy}, laser physics 
\cite{ikeda1,ikeda2,laser,hanna,gibbs}, etc. As is well known, a high
dimensional chaotic attractor characterizes these infinite dimensional
systems. It has been shown that the Lyapunov dimension of the attractor is
proportional to the characteristic delay time of the dynamics \cite%
{gibbs,farmer,procaccia,berreR}. Therefore, \textquotedblleft \textit{%
synchronization of hyperchaotic nonlinear delay dynamics}\textquotedblright\
has also been extensively explored from both a theoretical point of view as
well as a resource for communication schemes \cite{porte,lai,yinghai,rhodes}%
. A new aspect introduced in this case is the possibility of synchronizing
two chaotic dynamics with a time shift, giving rise to the phenomenon of
anticipated \cite%
{voss1,voss2,voss3,pyragas,masoller,zanette,mirasso,tang,toral,shore,heil}
(or retarded) synchronization, i.e., one of the chaotic systems (slave or
receiver system) follows the chaotic trajectory of the other one (master or
transmitter system) with an advanced (or retarded) time shift.

In any real experimental setup where chaotic synchronization is observed one
is naturally confronted with two undesirable effects that avoid reaching a
perfectly synchronized regime. The characteristic parameters of both systems
are not exactly the same, small \textit{parameter mismatch} may induce
clearly observable effects \cite{colet,mismatch,mar}. Also, departure with
respect to the perfectly synchronized manifold may also be due to intrinsic%
\textit{\ noise fluctuations} present in both systems \cite{tufillaro,chua}.
Both effects have been analyzed in the literature. Nevertheless, due to the
chaotic character of the dynamics, in general it is hard to obtain an
analytical estimation of these effects, which in fact may depend on the
specific nature of the chaotic systems as well as on the coupling scheme
used to achieve synchronization.

The aim of this paper is to provide a simple analytical description of the
phenomenon of synchronization of hyperchaotic delay dynamics, considering
realistic situations such as the presence of parameter mismatch as well as
the presence of intrinsic noise fluctuations in both synchronized systems.
We demonstrate that this goal can be achieved when the synchronized dynamics
are in a \textit{fully-developed chaos regime} \cite{gyorgyi}. In this
situation, the corresponding chaotic attractor does not have any stable
periodic orbit and its basin of attraction fills out almost the whole
available domain. These properties suggest that the nonlinear delay
contribution terms of the chaotic dynamics may be statistically equivalent
to an ergodic noise source. As demonstrated in Ref.~\cite{berre,berre1},
this property, in a long time limit and depending on the characteristic
parameter values, is in fact valid for a broad class of scalar delay
dynamics. Therefore, our \textit{main idea} consists in replacing the full
set of coupled delay chaotic evolutions that lead to synchronization by a
set of \textit{correlated Langevin evolutions} obtained from the original
chaotic ones after replacing the nonlinear delay contributions by a noise
term. Since the final Langevin equations are linear, their statistical
properties can be obtained in an exact analytical way, providing a simple
framework for characterizing the synchronization phenomenon. Departure with
respect to perfect synchronization is characterized in terms of a \textit{%
similarity function} \cite{kurths}, which measures the degree of correlation
between the quasi-synchronized dynamics.

The paper is organized as follows. In Sec. II we review the conditions under
which hyperchaotic delay dynamics can be represented by a Langevin dynamics.
In Sec. III we analyze the phenomenon of anticipated synchronization when
perturbed by external additive noises. In Sec. IV we analyze the effect of
parameters mismatch in a hyperchaotic communication scheme \cite{porte}. In
Sec. V we study a set of second order differential delay equations
associated to an electro-optical laser device \cite{colet} under the effect
of parameter mismatch and under the action of external additive noises. In
all cases, we present numerical simulations that sustain our theoretical
results. In Sec. VI we give the conclusions.

\section{Langevin approach to hyperchaotic delay dynamics in the
fully-developed chaos regime}

We will consider scalar nonlinear delay evolutions with the structure%
\begin{equation}
\dot{u}(t)=-\gamma u(t)+\beta f[u(t-T)]+\mathrm{I}(t).  \label{uEvolution}
\end{equation}%
Here, $\gamma $ defines a dissipative time scale and $T$ is the
characteristic time delay. The parameter $\beta $ controls the weights of
the nonlinear function $f(x),$ which is assumed to be \textit{oscillatory}
or at least exhibiting many different extrema \cite{notaReferee}. The term $%
\mathrm{I}(t)$ represents an extra inhomogeneous contribution that may
corresponds to a stationary Gaussian white noise (Sec. III), a deterministic
signal (Sec. IV) or even it may be a linear functional of the process $u(t)$
(Sec. V).

The parameter $\beta T$ may be considered as a complexity control parameter.
In fact, when $\beta T\gg 1,$ the dynamic reaches a fully-developed chaos
regime, where the stationary statistical properties of the corresponding
attractor can be reproduced with a linear Langevin equation \cite%
{berre,berre1}. This property can be understood by integrating Eq.~(\ref%
{uEvolution}), in \textit{the long time limit} $(\gamma t\gg 1),$ as%
\begin{equation}
\tilde{u}(t)\approx \int_{0}^{t}dt^{\prime }e^{-\gamma (t-t^{\prime
})}f[\beta \tilde{u}(t^{\prime }-T)+\mathrm{\tilde{I}}(t^{\prime }-T)],
\label{argument}
\end{equation}%
where $\tilde{u}(t)=[u(t)-\mathrm{\tilde{I}}(t)]/\beta ,$ and $\mathrm{%
\tilde{I}}(t)$\ is defined by $\mathrm{\tilde{I}}(t)\equiv
\int_{0}^{t}dt^{\prime }e^{-\gamma (t-t^{\prime })}\mathrm{I}(t^{\prime }).$
Then, by writing the integral operation over $f(x)$ as a discrete sum, $%
\int_{0}^{t}dt^{\prime }g(t^{\prime })\rightarrow \sum_{j}dt^{\prime
}g(jdt^{\prime }),$ one realizes that $\tilde{u}(t)$ may be considered as
the result of the addition of many \textquotedblleft \textit{statistically
independent}\textquotedblright contributions. This last property follows
from the fact that $f[\beta x]$ oscillates so fast that it behaves as a
driving random force. The characteristic correlation time [in units of time $%
1/\gamma $] of the nonlinear force $f[\beta x]$ is of order $\gamma /\beta $ 
\cite{berre,berre1}. When the correlation time is the small time-scale of
the problem, i.e., much shorter than the characteristic delay time, $\gamma
/\beta \ll \gamma T,$ and consistently much shorter than the characteristic
dissipative time, $\gamma /\beta \ll 1,$ the statistical independence of the
different contributions follows. Notice that this last property is
independent of the structure\ of the inhomogeneous term $\mathrm{\tilde{I}}%
(t),$ which only introduce a shift in the argument of $f[\beta x].$

From the previous analysis, in the parameters region $\beta T\gg 1$ and $\gamma /\beta \ll 1,$  
the central limit theorem \cite{kampen} tell us that 
$\tilde{u}(t)$ is a Gaussian process. Due to this characteristic, this regime
is also named \textit{Gaussian chaos}. Only when $\mathrm{\tilde{I}}(t-T)$
is a non-linear function (functional) of $u(t)$ \cite{berre1}, the long time
statistic may depart from a Gaussian one.

The lack of correlation between the different contributions of the
\textquotedblleft chaotic force\textquotedblright\ $f[\beta \tilde{u}(t-T)+%
\mathrm{\tilde{I}}(t-T)],$ allows to introduce the following Ansatz. In the
fully-developed chaos regime, \textquotedblleft \textit{the long time
statistical properties}\textquotedblright\ of $u(t)$ can be equivalently
obtained from Eq.~(\ref{argument}) after replacing the chaotic force by a
noise contribution $(\beta f[\tilde{u}(t-T)+\mathrm{\tilde{I}}%
(t-T)]\rightarrow \eta (t-T)),$ delivering the Langevin equation%
\begin{equation}
\dot{u}(t)=-\gamma u(t)+\eta (t-T)+\mathrm{I}(t).  \label{LangevinU}
\end{equation}
The noise $\eta (t)$ must have the same statistical properties as the
delayed chaotic force. Since we are restricting our analysis to the Gaussian
chaos regime, it is sufficient to map the first two statistical moments \cite%
{Markov}%
\begin{eqnarray}
\overline{\eta (t)} &=&\lim_{t\rightarrow \infty }\beta \overline{f[u(t)]},
\label{maping1} \\
\overline{\eta (t+\tau )\eta (t)} &=&\lim_{t\rightarrow \infty }\beta ^{2}%
\overline{f[u(t+\tau )]f[u(t)]}.  \label{maping2}
\end{eqnarray}%
The limit operation ($\lim_{t\rightarrow \infty }$) is introduced because
the statistical mapping is only valid in the long time regime. The overbar
symbol denotes an average over realizations obtained from Eq.~(\ref%
{uEvolution}) with different initial conditions. Equivalently, since the
chaotic attractor is ergodic (by definition of fully-developed chaos regime)
when the inhomogeneous term is statistically stationary \cite{kampen}, this
average can also be considered as a time average. For example, $%
\lim_{t\rightarrow \infty }\overline{f[u(t)]}=\lim_{t\rightarrow \infty
}(1/t)\int_{0}^{t}dt^{\prime }f[u(t^{\prime })].$

In agreement with the lack of correlation between the different
contributions of the chaotic force, $\eta (t)$ can be approximated by a
delta-correlated noise,%
\begin{equation}
\overline{\eta (t)\eta (t^{\prime })}-\overline{\eta (t)}\ \overline{\eta
(t^{\prime })}=\mathcal{A}\delta (t-t^{\prime }).  \label{D_Correlation}
\end{equation}%
This white noise approximation applies for time intervals $(t-t^{\prime })$
larger than the characteristic time correlation $(\gamma /\beta )$ [in units
of time $1/\gamma $] of the chaotic force \cite{berre,berre1}, i.e., $\gamma
(t-t^{\prime })>(\gamma /\beta ).$

The coefficient $\mathcal{A}$ measures the amplitude of $f[u(t)].$ Clearly, $%
\mathcal{A}$ must be proportional to $\beta ^{2}.$ Nevertheless, its exact
value is not universal and depends on the specific function $f(x)$ \cite%
{berre} as well as on the parameters that define the inhomogeneous term (see
next Section).

When $\mathrm{I}(t)$ is defined by an external driving force, we will assume
that 
\begin{equation}
\overline{\lbrack \eta (t)-\overline{\eta (t)}]\mathrm{I}(t^{\prime })}=0,
\label{descorrelacion}
\end{equation}%
i.e., the noise fluctuations representing the chaotic force and the
inhomogeneous term are statistically uncorrelated. The plausibility of this
assumption follows from the fact that the white nature of $\eta (t)$ only
relies on the rapid oscillating nature of $f[\beta x]$ while it is not
affected by the shift introduced by the functional $\mathrm{\tilde{I}}(t).$
Under this condition, the inhomogeneous contribution
only affects the mean value of the Gaussian profile associated to $u(t).$

Finally, we will assume that $\overline{\eta (t)}=0.$ The validity of this
condition only depends on the specific properties of the non-linear chaotic force.
In fact, the rapid oscillating nature of $f[\beta x]$
allows to discarding the asymmetry introduced by $\mathrm{\tilde{I}}(t)$
[Eq.(\ref{argument})]. As will become clear in the next section, the
generalization to the case $\lim_{t\rightarrow \infty }\beta \overline{%
f[u(t)]}\neq 0$ can also be worked straightforwardly.

With the noise correlation Eq.~(\ref{D_Correlation}), the stochastic
evolution Eq.~(\ref{LangevinU}) becomes a (driven) Orstein-Ulenbeck process 
\cite{kampen}. The statistical equivalence in the stationary and fully
developed chaos regimes of the deterministic evolution Eq.~(\ref{uEvolution}%
) and the Langevin Eq.~(\ref{LangevinU}), (without the inhomogeneous term)
was proved in Refs.~\cite{berre,berre1}. Clearly, this stochastic
representation does not provide any new information about the chaotic
dynamics. Nevertheless, in the next sections we will use this equivalence
for formulating a simple framework that allows us to get an analytical
characterization of the chaotic synchronization phenomenon for different
realistic circumstances.

\section{Anticipated synchronization subject to external additive noises}

In this section we will apply the previous Langevin representation of a
hyperchaotic attractor to analyze the phenomenon of anticipated
synchronization \cite{voss1,voss2,voss3} in the presence of external noise
sources. We consider a complete replacement scheme \cite{toral}, defined by
the \textit{coupled chaotic evolutions} 
\begin{subequations}
\label{MasterSlave}
\begin{eqnarray}
\dot{x}(t) &=&-\gamma x(t)+\beta f[x(t-T)]+\xi _{x}(t),
\label{MaestroCaotico} \\
\dot{y}(t) &=&-\gamma y(t)+\beta f[x(t)]+\xi _{y}(t).
\end{eqnarray}%
In this context, the variables $x(t)$ and $y(t)$ are referred as master and
slave variables respectively. As before, $\gamma $ is a constant dissipative
rate, and $\beta $ measures the weight of the delay-nonlinear contribution $%
f(x).$

We have considered external noise sources, defined by the master and slave
noises $\xi _{x}(t)$ and $\xi _{y}(t)$ respectively. We assume that their
mean values are null $\left\langle \xi _{z}(t)\right\rangle =0,$ where $%
z=x,y,$ and $\left\langle \cdots \right\rangle $ denotes average over noise
realizations. Furthermore, we assume that both noises are Gaussian, with
correlations 
\end{subequations}
\begin{equation}
\left\langle \xi _{z}(t)\xi _{z^{\prime }}(t^{\prime })\right\rangle =%
\mathcal{A}_{zz^{\prime }}\delta (t-t^{\prime }).
\label{ExternalNoiseCorrelation}
\end{equation}%
The \textquotedblleft diffusion\textquotedblright\ coefficients\ satisfy the
positivity constraint $\mathcal{A}_{xx}\mathcal{A}_{yy}-\mathcal{A}_{xy}%
\mathcal{A}_{yx}\geq 0,$ $(\mathcal{A}_{xy}=\mathcal{A}_{yx})$ \cite{kampen}%
. These definitions allows to cover the case where both the master and slave
variables are affected by intrinsic uncorrelated fluctuations, $\mathcal{A}%
_{xy}=0,$ as well as the case of correlated fluctuations $\mathcal{A}_{xy}>0.
$ This last situation may be easily produced in any experimental setup. In
particular, for $\mathcal{A}_{xx}=\mathcal{A}_{yy}=\mathcal{A}_{xy},$ the
noises that drive the master and slave dynamics are exactly the same, i.e.,
Eq.~(\ref{MasterSlave}) with $\xi _{x}(t)=\xi _{y}(t).$ This property
follows by diagonalizing the matrix of diffusion noise coefficients $\{\{%
\mathcal{A}_{xx},\mathcal{A}_{xy}\},\{\mathcal{A}_{xy},\mathcal{A}_{yy}\}\}.$

As it is well known, in the absence of the external noises $\xi _{x}(t)$ and 
$\xi _{y}(t),$ the master-slave dynamics Eq.~(\ref{MasterSlave}), after a
time transient of order $1/\gamma ,$ reach the synchronized manifold $%
x(t+T)=y(t).$ Therefore, the slave variable \textit{anticipates} the
behavior of the master variable. The achievement of this state is clearly
affected by the presence of the external noises. The degree of departure
with respect to the perfectly synchronized manifold can be measured with a 
\textit{similarity function}, defined as \cite{kurths}%
\begin{equation}
S(\tau )\equiv \lim_{t\rightarrow \infty }\left[ \frac{\langle \overline{%
[x(t+\tau )-y(t)]^{2}}\rangle }{[\langle \overline{x^{2}(t)}\rangle \langle 
\overline{y^{2}(t)}\rangle ]^{1/2}}\right] ^{1/2}.  \label{Similaridad}
\end{equation}%
As before, the overbar denotes an average with respect to the system
initialization or equivalently a time average in the asymptotic regime. In
the absence of the external noises, this object satisfies $S(T)=0,$
indicating that the perfectly synchronized state $x(t+T)=y(t)$ was achieved.
In the presence of the noises, we expect $S(T)>0.$

The characterization of the behavior of the similarity function $S(\tau )$
from the evolution Eq.~(\ref{MasterSlave}) is in principle \textit{a highly
non trivial task}. The major complication come from the chaotic nature of
the master and slave dynamics. Even in the absence of the external noises,
in general it is impossible to get an analytical expression for the
similarity function. Nevertheless, if both dynamics are in the
fully-developed chaos regime, from the previous section, we know that a
simpler representation may be achieved. The Langevin approach to Eq.~(\ref%
{MasterSlave}) reads 
\begin{subequations}
\label{NoiseAprox}
\begin{eqnarray}
\dot{x}(t) &=&-\gamma x(t)+\eta (t-T)+\xi _{x}(t),  \label{NoiseAproxUNO} \\
\dot{y}(t) &=&-\gamma y(t)+\eta (t)+\xi _{y}(t).  \label{NoiseAproxDOS}
\end{eqnarray}%
Notice that this equation corresponds to Eq.~(\ref{MasterSlave}) with the
replacement $f[x]\rightarrow \eta $ and maintaining the respective time
arguments. As before, the effective noise $\eta (t)$ is defined by the
correlation Eq.~(\ref{D_Correlation}). Furthermore, we will assume that $%
\overline{\eta (t)}=0.$ We will deal the case, $\overline{\eta (t)}\neq 0$
at the end of this section.

While the nature of Eqs.~(\ref{NoiseAprox}) is completely different to that
of Eqs.~(\ref{MasterSlave}), in the asymptotic time regime these Langevin
equations, without the external noises $\xi _{x}(t)$ and $\xi _{y}(t),$ also
reach a perfectly synchronized state. In fact, without the external noises,
from Eq.~(\ref{NoiseAprox}) it is possible to write $(d/dt)[x(t)-y(t-T)]=-%
\gamma \lbrack x(t)-y(t-T)],$ implying that after a time transient of order $%
(1/\gamma )$ the manifold $x(t+T)=y(t)$ is reached. From the previous
section we know that the statistical properties of the corresponding master
process Eq.~(\ref{MaestroCaotico}) and Eq.~(\ref{NoiseAproxUNO}) are the
same. Then, we estimate the similarity function Eq.~(\ref{Similaridad})
associated to the nonlinear evolution Eq.~(\ref{MasterSlave}) from the
simpler linear Langevin evolutions Eq.~(\ref{NoiseAprox}).

In the long time limit $(\gamma t\gg 1),$ the master and slave Langevin
evolutions can be integrated for each realization of the noise $\eta (t),$
which represent the nonlinear force, and external noises $[\xi _{x}(t)$ and $%
\xi _{y}(t)]$ as $x(t)\approx \int_{0}^{t}dt^{\prime }e^{-\gamma
(t-t^{\prime })}[\eta (t^{\prime }-T)+\xi _{x}(t^{\prime })],$ and as $%
y(t)\approx \int_{0}^{t}dt^{\prime }e^{-\gamma (t-t^{\prime })}[\eta
(t^{\prime })+\xi _{y}(t^{\prime })],$ respectively. By using the
correlations Eq.~(\ref{D_Correlation}) [with $\overline{\eta (t)}=0$] and
Eq.~(\ref{ExternalNoiseCorrelation}), it follows 
\end{subequations}
\begin{equation}
\lim_{t\rightarrow \infty }\langle \overline{x^{2}(t)}\rangle =\frac{%
\mathcal{A}+\mathcal{A}_{xx}}{2\gamma },\ \ \ \ \ \ \lim_{t\rightarrow
\infty }\langle \overline{y^{2}(t)}\rangle =\frac{\mathcal{A}+\mathcal{A}%
_{yy}}{2\gamma }.  \label{CuadraticAverageMS}
\end{equation}%
In a similar way, we get%
\begin{equation*}
\lim_{t\rightarrow \infty }\langle \overline{x(t+\tau )y(t)}\rangle \!=\!%
\frac{\mathcal{A}}{2\gamma }\exp [-\gamma |\tau -T|]\!+\!\frac{\mathcal{A}%
_{xy}}{2\gamma }\exp [-\gamma |\tau |].
\end{equation*}%
Therefore, the similarity function reads%
\begin{equation}
S(\tau )\!=\!\sqrt{2}\!\left[ \!\frac{(1+\frac{\mathcal{A}_{xx}+\mathcal{A}%
_{yy}}{2\mathcal{A}})-(e^{-\gamma |\tau -T|}+\frac{\mathcal{A}_{xy}}{%
\mathcal{A}}e^{-\gamma |\tau |})}{(1+\frac{\mathcal{A}_{xx}}{\mathcal{A}}%
)^{1/2}(1+\frac{\mathcal{A}_{yy}}{\mathcal{A}})^{1/2}}\!\right] ^{1/2}\!.
\label{SimilNoiseIkeda}
\end{equation}%
This is one of the main results of this section. Notice that this expression
only depends on one free parameter, i.e., the amplitude of the chaotic force 
$\mathcal{A}.$ On the other hand, this result relies in assuming the absence
of any statistical correlation between the noise $\eta (t)$ representing the
chaotic force and the external noises $\{\xi _{x}(t),\xi _{y}(t)\}.$
Consistently with the delta correlated nature of both contributions, Eqs.~(%
\ref{D_Correlation}) and (\ref{ExternalNoiseCorrelation}), we have also
calculated the extra contributions to Eq.~(\ref{SimilNoiseIkeda}) that
appear by assuming a delta cross correlation between both kind of objects.
Nevertheless, the numerical simulations presented along the paper contradict
the existence of any extra correlation, supporting the (previous) arguments
that explain the statistical independence between the chaotic force and any
external source, Eq.~(\ref{descorrelacion}).

From Eq.~(\ref{SimilNoiseIkeda}) one can analyze different limits. In the
absence of external noise we get%
\begin{equation}
S(\tau )=\sqrt{2}(1-\exp [-\gamma |\tau -T|])^{1/2}.
\label{SimilIkedaDeterm}
\end{equation}%
Note that this expression does not depends on the chaotic force amplitude$%
\mathcal{\ A},$ it is defined only in terms of the local dissipation rate $%
\gamma $ and the delay $T.$ Consistently, $S(\tau )$ satisfies the
anticipated synchronization condition $S(T)=0.$ When the external noise
sources are taken in account, we get%
\begin{equation}
S(T)=\sqrt{2}\left[ \frac{\frac{\mathcal{A}_{xx}+\mathcal{A}_{yy}}{2\mathcal{%
A}}-\frac{\mathcal{A}_{xy}}{\mathcal{A}}e^{-\gamma T}}{(1+\frac{\mathcal{A}%
_{xx}}{\mathcal{A}})^{1/2}(1+\frac{\mathcal{A}_{yy}}{\mathcal{A}})^{1/2}}%
\right] ^{1/2}.
\end{equation}%
This value measures the departure with respect to the perfectly synchronized
manifold $x(t+T)=y(t).$ Notice that the correlation $\mathcal{A}_{xy}$
always decrease the value of $S(T).$ This effect is exponentially diminished
when increasing $\gamma T.$

\subsection{Anticipated synchronization of delay maps}

The previous results can be extended to the case of anticipated
synchronization in discrete delay maps. We consider the maps that follow
after discretizing the time variable in Eq.~(\ref{MasterSlave}), $t=n\delta
t,$ and integrating both the master and slave evolutions up to first order
in the discrete time step $\delta t.$ We get 
\begin{subequations}
\label{NoiseDrivenMap}
\begin{eqnarray}
x_{n+1} &=&ax_{n}+bf(x_{n-n_{0}})+\xi _{n}^{x}, \\
y_{n+1} &=&ay_{n}+bf(x_{n})+\xi _{n}^{y}.
\end{eqnarray}%
Here, $n_{0}$ defines the characteristic delay step and the
\textquotedblleft dissipative\textquotedblright\ rate satisfies $0<a<1.$ For
each $n,$ $\xi _{n}^{x}$ and $\xi _{n}^{y},$ are independent Gaussian
distributed variables with $\langle \xi _{n}^{z}\xi _{m}^{z^{\prime
}}\rangle =A_{zz^{\prime }}\delta _{nm},$ $(z=x,y),$ subject to the
constraint $A_{xx}A_{yy}-A_{xy}A_{yx}\geq 0,$ $(A_{xy}=A_{yx})$ \cite{kampen}%
. The parameters of the continuous time evolution Eq.~(\ref{MasterSlave})
and the discrete map Eq.~(\ref{NoiseDrivenMap}) are related by
\end{subequations}
\begin{equation}
\gamma =\frac{1-a}{\delta t},\ \ \ \ \beta =\frac{b}{\delta t},\ \ \ \
T=n_{0}\delta t,\ \ \ \ \mathcal{A}_{zz^{\prime }}=\frac{A_{zz^{\prime }}}{%
\delta t}.  \label{mapping}
\end{equation}%
For the discrete map, the similarity function Eq.~(\ref{Similaridad}) is
defined as%
\begin{equation}
S_{m}=\lim_{n\rightarrow \infty }\left[ \frac{\langle \overline{%
(x_{n+m}-y_{n})^{2}}\rangle }{\sqrt{\langle \overline{x_{n}^{2}}\rangle \
\langle \overline{y_{n}^{2}}\rangle }}\right] ^{1/2}.
\label{SimilarDiscreta}
\end{equation}%
As we will show in the next examples, this object can be fitted by Eq.~(\ref%
{SimilNoiseIkeda}) with the mapping Eq.~(\ref{mapping}). This property is
valid when the discrete map provides a good approximation to the continuous
time evolution. In fact, when $a\approx 1,$ i.e., $\gamma \delta t\ll 1,$
the map Eq.~(\ref{NoiseDrivenMap}) can be read as a numerical algorithm for
simulating the continuous time evolution Eq.~(\ref{NoiseAprox}).
Nevertheless, we remark that Eq.~(\ref{NoiseDrivenMap}) can also be analyzed
without appealing to the parameter mapping Eq.~(\ref{mapping}), i.e., the
Langevin approach can also be formulated for discrete time dynamic. $S_{m}$
can be estimated after replacing the chaotic force in Eq.~(\ref%
{NoiseDrivenMap}) by a discrete noise, $f(x)\rightarrow \eta _{n},$ with $%
\overline{\eta _{n}}=0$ and $\overline{\eta _{n}\eta _{m}}=A\delta _{nm}.$
We get%
\begin{equation}
\lim_{n\rightarrow \infty }\langle \overline{x_{n}^{2}}\rangle =\frac{%
A+A_{xx}}{1-a^{2}},\ \ \ \ \ \ \ \ \lim_{n\rightarrow \infty }\langle 
\overline{y_{n}^{2}}\rangle =\frac{A+A_{yy}}{1-a^{2}}.
\label{CuadraticMSmap}
\end{equation}%
The master-slave correlation reads%
\begin{equation}
\lim_{n\rightarrow \infty }\langle \overline{(x_{n+m}y_{n})}\rangle \!=\frac{%
1}{1-a^{2}}(Aa^{|m-no|}+A_{xy}a^{|m|}).  \label{XYCorrelationMAP}
\end{equation}%
Then, the (discrete) similarity function reads%
\begin{equation}
S_{m}=\sqrt{2}\left[ \frac{(1+\frac{A_{xx}+A_{yy}}{2A})-(a^{|m-no|}+\frac{%
A_{xy}}{A}a^{|m|})}{(1+\frac{A_{xx}}{A})^{1/2}(1+\frac{A_{yy}}{A})^{1/2}}%
\right] ^{1/2}.  \label{SimildiscretaIkeda}
\end{equation}%
Consistently, by using Eq.~(\ref{mapping}) and the mapping between the
chaotic force amplitudes%
\begin{equation}
\mathcal{A}=A\delta t^{-1},
\end{equation}%
from Eq.~(\ref{SimildiscretaIkeda}) to first order in $\delta t,$ one
recovers Eq.~(\ref{SimilNoiseIkeda}). Furthermore, the conditions that
guarantee the validity of the Langevin representation in the continuous time
case, i.e., $\beta T\gg 1$ and $\gamma /\beta \ll 1,$ from the mapping Eqs. (%
\ref{mapping}), here read $bn_{0}\gg 1$ and $(1-a)/b\ll 1.$

\subsection{Numerical results}

To check the validity of the previous results, we consider an\ Ikeda-like
delay-differential equation \cite{ikeda1,ikeda2}, i.e. Eq.~(\ref{MasterSlave}%
) with $f[x]=\sin (x).$ Its associated discrete map [Eq.~(\ref%
{NoiseDrivenMap})] read
\begin{subequations}
\label{IKEDOL_MAPA}
\begin{eqnarray}
x_{n+1} &=&ax_{n}+b\sin (x_{n-n_{0}})+\xi _{n}^{x}, \\
y_{n+1} &=&ay_{n}+b\sin (x_{n})+\xi _{n}^{y}.
\end{eqnarray}%
To obtain the following results, we generate a set of realizations from the
map Eq.~(\ref{IKEDOL_MAPA}) by considering different random initials
conditions in the interval $(-\pi ,\pi )$ for both, the master and the slave
variables. By averaging over this set of realizations $(\approx 10^{4}),$ we
determine numerically the similarity function Eq.~(\ref{SimilarDiscreta}).
To check the ergodic property of the corresponding chaotic attractor, we
repeated the numerical calculations by averaging over time a single
trajectory with an arbitrary set of initial conditions. Consistently, we
obtained the same results and characteristic behaviors.%
\begin{figure}[tb]
\includegraphics[height=6cm,bb=14 15 296 223]{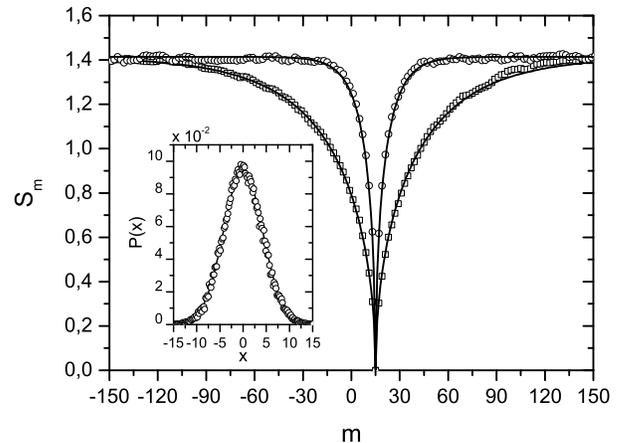}
\caption{Similarity function of the coupled delay maps Eq.~(\protect\ref%
{IKEDOL_MAPA}) without noise, $A_{xx}=A_{yy}=A_{xy}=0.$ The parameters are, $%
n_{0}=15,$ $b=3,$ and $a=0.975$ (squares) and $a=0.9$ (circles). The lines
correspond to the fitting Eq.~(\protect\ref{SimilIkedaDeterm}) joint with
the mapping Eq.~(\protect\ref{mapping}) [with $\protect\delta t=1],$
indistinguishable from Eq.~(\protect\ref{SimildiscretaIkeda}). The inset
corresponds to the stationary probability distribution $P(x)$ of the process 
$x_{n}$ (with $a=0.9$).}
\end{figure}

In Fig. 1 we show the similarity function when the external noises are
absent, i.e., the similarity corresponding to the deterministic map. In
agreement with this condition, we notice that $S_{n_{0}}=0,$ implying that
the manifold $x_{m+n_{0}}=y_{m}$ characterizes the asymptotic behavior.
Furthermore, we find that both the expression for the continuous time case
[Eq.~(\ref{SimilNoiseIkeda}) with the mapping Eq.~(\ref{mapping})], as well
as the similarity function of the map [Eq.~(\ref{SimildiscretaIkeda})] are
indistinguishable from each other (in the scale of the graphic) and both\
correctly fit the numerical behavior. In the inset, we show the stationary
probability distribution of the master process $x_{n}.$ In agreement with
our considerations, this distribution can be fit with a Gaussian
distribution. From its width, and by using Eq.~(\ref{CuadraticMSmap}) [or
Eq.~(\ref{CuadraticAverageMS})] we estimated the value of the chaotic force
amplitude, $A\approx 3.5.$

In the absence of the noises, we corroborate that by taking the function $%
f[x]=\sin (x+\phi ),$ independently of the value of the phase $\phi ,$ the
same statistical behaviors follow. This result confirms the arguments
presented in the previous section [Eq.~(\ref{argument})] about the
statistical invariance of the chaotic force under a shift of its argument.

In Fig. 2 we show the similarity function when the master and slave dynamics
are affected by two uncorrelated external noises. As expected, we found that 
$S_{n_{0}}>0$ $(n_{0}=15).$ In the inset, we show a characteristic
master-slave realization. Even in the presence of the external noises, both
trajectories are approximately the same (the slave anticipate the master
trajectory). 
\begin{figure}[t]
\includegraphics[height=6cm,bb=14 15 296 227]{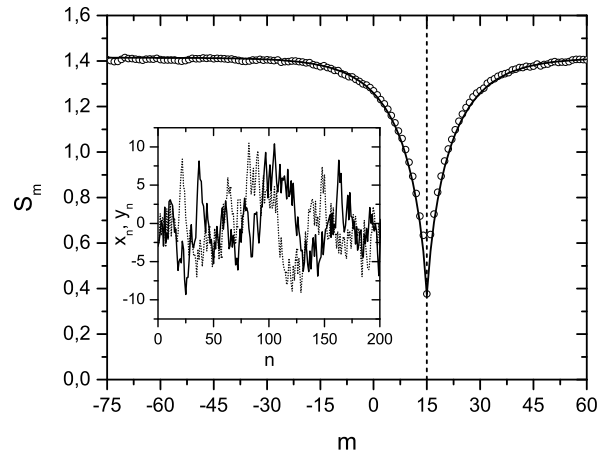}
\caption{Similarity function of the coupled delay maps Eq.~(\protect\ref%
{IKEDOL_MAPA}) subject to external Gaussian noise fluctuations. The
parameters are $n_{0}=15,$ $a=0.9,$ and $b=3.$ The noise parameters are $%
A_{xx}=A_{yy}=0.25$ and $A_{xy}=0.$ The line corresponds to the fitting Eq.~(%
\protect\ref{SimilIkedaDeterm}) joint with the mapping Eq.~(\protect\ref%
{mapping}) [with $\protect\delta t=1],$ indistinguishable from Eq.~(\protect
\ref{SimildiscretaIkeda}). The chaotic noise amplitude results $A\approx 3.5.
$ The inset corresponds to a master ($x_{n},$ full line) and slave ($y_{n},$
dotted line) realization.}
\end{figure}
In contrast with the previous figure, here the fitting to the similarity
function depends explicitly on the chaotic force amplitude $A.$ We found
that the value of $A$ that provides the best fitting is consistent with the
one found for the deterministic map (Fig. 1), i.e., $A\approx 3.5.$ Then, in
this case, the inequality $A\gg \{A_{xx},A_{yy}\}$ is satisfied, implying
that the fluctuations induced by the determinist chaotic dynamic are much
larger than those induced by the external noise sources.

Maintaining all the parameter values corresponding to Fig. 2, we analyzed
the case $A_{xx}=A_{yy}=A_{xy}.$ In this situation, the noises that drive
the master and slave dynamics are exactly the same, i.e., Eq.~(\ref%
{IKEDOL_MAPA}) with $\xi _{n}^{x}=\xi _{n}^{y}$ (as in the continuous time
case, this property follows by diagonalizing the matrix of diffusion noise
coefficients $\{\{A_{xx},A_{xy}\},\{A_{xy},A_{yy}\}\}$). We found that the
similarity function is almost indistinguishable from that of Fig. 2.
Independently of the external noise correlations, in both cases our approach
provides a very good fitting of the numerical results.

In Fig. 3, by maintaining the parameters of the deterministic map, i.e., $%
(a,b,n_{0}),$ we increased the amplitude of the external noises, such that $%
\{A_{xx},A_{yy}\}\approx A.$ Then, the external noise-induced fluctuations
are of the same order as the intrinsic chaotic dynamical fluctuations. Even
in this limit, our approach provides an excellent fitting of the numerical
results. Both, the case of uncorrelated noises ($A_{xy}=0$ with $%
A_{xx}=A_{yy}$) and the case of completely correlated noises ($%
A_{xx}=A_{yy}=A_{xy}$) are considered. In both situations, since the
external noise amplitudes are larger than in Fig. 2, the value of $S_{n_{0}}$
increases, indicating a weaker (anticipated) synchronization between the
master and slave dynamics. 
\begin{figure}[tb]
\includegraphics[height=6cm,bb=14 15 296 227]{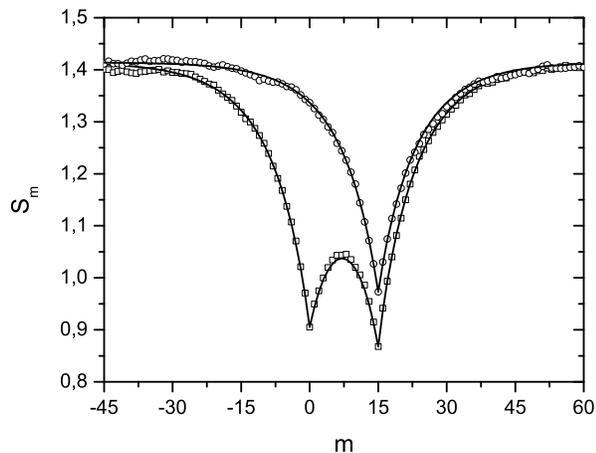}
\caption{Similarity function of the coupled delay maps Eq.~(\protect\ref%
{IKEDOL_MAPA}) driven by different noises. The circles correspond to $%
A_{xx}=A_{yy}=3.5$ and $A_{xy}=0,$ while the squares correspond to $%
A_{xx}=A_{yy}=A_{xy}=3.5.$ The parameters of the map are the same than in
Fig. 2. The lines correspond to the fitting Eq.~(\protect\ref%
{SimilIkedaDeterm}) joint with the mapping Eq.~(\protect\ref{mapping}) [with 
$\protect\delta t=1],$ indistinguishable from Eq.~(\protect\ref%
{SimildiscretaIkeda}). The chaotic noise amplitude results $A\approx 4.3.$}
\end{figure}

In the case of completely correlated noises, $A_{xx}=A_{yy}=A_{xy},$ the
similarity function develop two (local) minima, one at $m=n_{0}$ and other
at $m=0.$ This feature follows from the competition between two different
synchronizing mechanisms, induced by the chaotic dynamics and the external
noises, respectively. In fact, as the noise that drives the master and slave
dynamics is the same, its action tends to synchronize both dynamics without
any time shift [see Eq.~(\ref{IKEDOL_MAPA}) with $\xi _{n}^{x}=\xi _{n}^{y}$%
], producing the dip at $m=0.$ In agreement with this argument, from Eq.~(%
\ref{SimildiscretaIkeda}) [or equivalently Eq.~(\ref{SimilNoiseIkeda})] one
can deduce that only one minimum at $m=0$ will be appreciable in the
similarity function when $A\ll A_{xx}=A_{yy}=A_{xy},$ i.e., in the limit of
high noise intensity.

As in Fig. 2, the similarity function depends on the amplitude $A$ of the
chaotic force. Here, the best fitting to the similarity function is obtained
with $A\approx 4.3.$ This value is larger than the one associated to the
deterministic dynamics [Fig. 1] or the case of weak external noises [Fig.
2]. Then, in general it is necessary to consider that the chaotic force
amplitude $A$ is also a function of the external noise intensities, $%
A=A[\{f(x),b\},A_{xx},A_{yy})].$ We remark that in concordance with the
Langevin representation, the value of $A$ determined from Eqs.~(\ref%
{CuadraticMSmap}) to (\ref{SimildiscretaIkeda}) is the same. By using the
parameters of Fig. 3, we found a moderate dependence on the external noise
intensities, i.e., the maximal variation of $A$ with the amplitude of the
noises does not exceeds thirty percent (30\%) of the deterministic map value
(Fig. 1, $A\approx 3.5$). The dependence is smooth but non-monotonous. We
found that $A$ saturates to a fixed value ($A\approx 4.5$) when increasing
the external noise amplitudes, $A<\{A_{xx},A_{yy}\}.$

\subsection{Chaotic force with a non-null average value}

Our previous theoretical calculations rely on the assumption that the
average (over realizations or its stationary time average) of the chaotic
force is zero, i.e., $\overline{f[x(t)]}=0.$ When the dynamic, that does not
include the non-linear contribution, is purely dissipative [Eq. (\ref%
{uEvolution}) with $f(x)\rightarrow 0],$ the validity of this assumption
requires that $f(x)$ takes symmetrically positive and negative values.
Functions that do not satisfy this property also arise in real experimental
setups \cite{hanna,porte,colet}. In these cases, due to the pure dissipative
nature of the evolution Eq.~(\ref{MasterSlave}) [or Eq.~(\ref{NoiseDrivenMap}%
)], in the long time limit the master-slave realizations will fluctuate
around a non-null fixed value.

This situation can be managed by writing $f(x)=[f(x)-\overline{f(x)}]+%
\overline{f(x)}.$ Then, the previous theoretical calculations can be easily
extended by replacing $[f(x)-\overline{f(x)}]\rightarrow \eta (t)$ [with $%
\overline{\eta (t)}=0$] and by maintaining the extra contribution $\overline{%
f(x)}$ in the final Langevin representation. For example, taking the
functions $f(x)=\sin ^{2}(x)$ or $f(x)=\cos ^{2}(x)$ in Eq.~(\ref%
{MasterSlave}), from the Langevin representation Eq.~(\ref{NoiseAprox}), it
is possible to deduce that in the fully-developed chaos regime the
master-slave realizations will fluctuate around $\beta /(2\gamma ).$

When the dynamic without the non-linear contribution can by itself induce
symmetric dynamical oscillations, the symmetry requirement on the function $%
f(x)$ may be eliminated [see Sec.\ V].

\section{Effect of parameters mismatch in a hyperchaotic communication scheme%
}

Hyperchaotic delay dynamics may be used as a resource for secure encoded
communication. Different schemes have been proposed, in all the cases, it is
argued that the security of the communication channel may be improved by
increasing the dimension of the hyperchaotic attractor. Here, we study the
dynamics \cite{porte} 
\end{subequations}
\begin{subequations}
\label{Message}
\begin{eqnarray}
\dot{x}(t) &=&-\gamma x(t)+\beta f[x(t-T)]+\mathfrak{M}(t), \\
\dot{y}(t) &=&-\gamma ^{\prime }y(t)+\beta ^{\prime }f[x(t-T^{\prime })].
\end{eqnarray}%
The variable $x(t)$ is the carrier signal where the message is encoded. The
external feed $\mathfrak{M}(t)$ is defined by $\mathfrak{M}(t)=[d\mathfrak{m}%
(t)/dt+\gamma \mathfrak{m}(t)],$ where $\mathfrak{m}(t)$ is the message to
be transmitted. The variable $y(t)$ is the receiver. It is easy to
demonstrate that in the stationary regime the state $x(t)-y(t)=\mathfrak{m}%
(t)$ is reached, implying that the receiver is able to unmask the message
encoded in the hyperchaotic dynamics of $x(t)$. This condition is only
satisfied when the characteristic parameters of the transmitter and receiver
evolutions [Eq.~(\ref{Message})] are the same, i.e., $\gamma ^{\prime
}=\gamma ,$ $\beta ^{\prime }=\beta ,$ and $T^{\prime }=T$ \cite{nota}. In
any real experimental situation, it is expected that this condition is not
fulfilled, i.e., the decodification of the message is performed in the
presence of an unavoidable (finite) parameter mismatching. For simplicity,
in the present section we will not consider the action of any external noise
perturbation source.

In order to achieve a general characterization of the influence of the
parameters mismatch, we take $\mathfrak{M}(t)=0.$ As in the previous
section, we will use the similarity function $S(\tau )$ [Eq.~(\ref%
{Similaridad})] as a measure of the degree of synchronization between the
emitter and receiver variables.

When the transmitter dynamics is in the fully-developed chaos regime, we can
extend the Langevin approach to the present situation. In order to estimate
the similarity function, we replace the chaotic coupled evolution Eq.~(\ref%
{Message}) by the Langevin equations 
\end{subequations}
\begin{subequations}
\label{MismacthLangevin}
\begin{eqnarray}
\dot{x}(t) &=&-\gamma x(t)+\eta (t-T), \\
\dot{y}(t) &=&-\gamma ^{\prime }y(t)+\sqrt{\alpha }\eta (t-T^{\prime }),
\end{eqnarray}%
where the parameter $\alpha $ is defined by 
\end{subequations}
\begin{equation}
\alpha \equiv \left( \frac{\beta ^{\prime }}{\beta }\right) ^{2}.
\label{alfalfa}
\end{equation}%
As before, the noise $\eta (t)$ is characterized by $\overline{\eta (t)}=0,$
and $\overline{\eta (t)\eta (t^{\prime })}=\mathcal{A}\delta (t-t^{\prime })$
[Eq.~(\ref{D_Correlation})]. The similarity function $S(\tau )$ associated
to Eq.~(\ref{MismacthLangevin}), can be determine analytically by a
straightforward calculation. We get%
\begin{equation}
S(\tau )=\left[ \sqrt{\frac{\gamma ^{\prime }}{\alpha \gamma }}+\sqrt{\frac{%
\alpha \gamma }{\gamma ^{\prime }}}-\frac{4\sqrt{\gamma \gamma ^{\prime }}}{%
(\gamma +\gamma ^{\prime })}\exp [-\bar{\gamma}|\tau -\Delta |]\right]
^{1/2},  \label{SimilMismatchIkeda}
\end{equation}%
where we have defined the delay time mismatch 
\begin{equation}
\Delta \equiv T-T^{\prime },
\end{equation}%
and the rate $\bar{\gamma}$ is defined as 
\begin{equation}
\bar{\gamma}\equiv \left\{ 
\begin{array}{c}
\gamma \ \ \ for\ \ \ \tau >\Delta  \\ 
\gamma ^{\prime }\ \ \ for\ \ \ \tau <\Delta 
\end{array}%
\right. .
\end{equation}%
We notice that due to the normalization constants in the definition of the
similarity function Eq.~(\ref{Similaridad}), the final expression Eq.~(\ref%
{SimilMismatchIkeda}) does not depend explicitly on the chaotic force
amplitude $\mathcal{A}$. The similarity function only satisfies $S(0)=0,$
when the transmitter-receiver parameters are exactly the same. In this
situation, the manifold $x(t)=y(t)$ characterize the stationary regime.

\subsection{Parameter mismatch in discrete delay maps}

The previous results can be extended to\ the coupled maps obtained from the
evolution Eq.~(\ref{Message}) after a first order Euler integration, i.e., 
\begin{subequations}
\label{IkedaMapMismatch}
\begin{eqnarray}
x_{n+1} &=&ax_{n}+bf(x_{n-n_{0}}), \\
y_{n+1} &=&a^{\prime }y_{n}+b^{\prime }f(x_{n-n_{0}^{\prime }}).
\end{eqnarray}%
Here, the transmitter and receiver parameters of both, the continuous and
the discrete time evolutions, must be related as
\end{subequations}
\begin{subequations}
\label{MappingMismacht}
\begin{eqnarray}
\gamma  &=&\frac{1-a}{\delta t},\ \ \ \ \beta =\frac{b}{\delta t},\ \ \ \
T=n_{0}\delta t, \\
\gamma ^{\prime } &=&\frac{1-a^{\prime }}{\delta t},\ \ \ \ \beta ^{\prime }=%
\frac{b^{\prime }}{\delta t},\ \ \ \ T^{\prime }=n_{0}^{\prime }\delta t,
\end{eqnarray}%
where $\delta t$ is the characteristic discretization time step. For the
discrete maps, the similarity function Eq.~(\ref{SimilarDiscreta}) reads 
\end{subequations}
\begin{eqnarray}
S_{m} &=&\left[ \sqrt{\frac{(1-a^{\prime 2})}{\alpha (1-a^{2})}}+\sqrt{\frac{%
\alpha (1-a^{2})}{(1-a^{\prime 2})}}\right.   \notag \\
&&\left. -\frac{2\sqrt{(1-a^{2})(1-a^{\prime 2})}}{(1-aa^{\prime })}\bar{a}%
^{|m-m_{0}|}\right] ^{1/2},  \label{SimilMapMismacht}
\end{eqnarray}%
where we have defined%
\begin{equation}
m_{0}\equiv n_{0}-n_{0}^{\prime },\ \ \ \ \ \ \ \ \alpha =\left( \frac{%
b^{\prime }}{b}\right) ^{2},  \label{alfatal}
\end{equation}%
and the \textquotedblleft dissipative rate\textquotedblright 
\begin{equation}
\bar{a}\equiv \left\{ 
\begin{array}{c}
a\ \ \ for\ \ \ m>m_{0} \\ 
a^{\prime }\ \ \ for\ \ \ m<m_{0}%
\end{array}%
\right. .
\end{equation}%
Notice that under the mapping Eq.~(\ref{MappingMismacht}), the definition of
the parameter $\alpha $ in Eq.~(\ref{alfatal}) is consistent with Eq.~(\ref%
{alfalfa}).

\subsection{Numerical results}

Here we consider the coupled chaotic maps 
\begin{subequations}
\label{MismatchCoseno}
\begin{eqnarray}
x_{n+1} &=&ax_{n}+b\cos (x_{n-n_{0}}), \\
y_{n+1} &=&a^{\prime }y_{n}+b^{\prime }\cos (x_{n-n_{0}^{\prime }}).
\end{eqnarray}%
In the inset of Fig. 4, we show a characteristic realization of the emitter
and receiver variables. By averaging over different realizations $(\approx
10^{4}),$ we get the similarity function. We notice that due to the specific
values of the characteristic parameters, the receiver \textquotedblleft
synchronizes\textquotedblright\ with the \textquotedblleft
past\textquotedblright\ of the transmitter variable. In fact, $S_{m}$
attains it minimal value at $m_{0}=-5.$ Furthermore, due to the difference
in the parameters $b$ and $b^{\prime },$ the amplitude of the receiver
fluctuations are bigger than those of the transmitter.

In contrast with Fig. 2, here the similarity function is not symmetrical
around its minimal value. This asymmetry has its origin in the mismatch
between the \textquotedblleft dissipative rate\textquotedblright\ parameters 
$a$ and $a^{\prime }.$ 
\begin{figure}[t]
\includegraphics[height=6cm,bb=14 15 296 227]{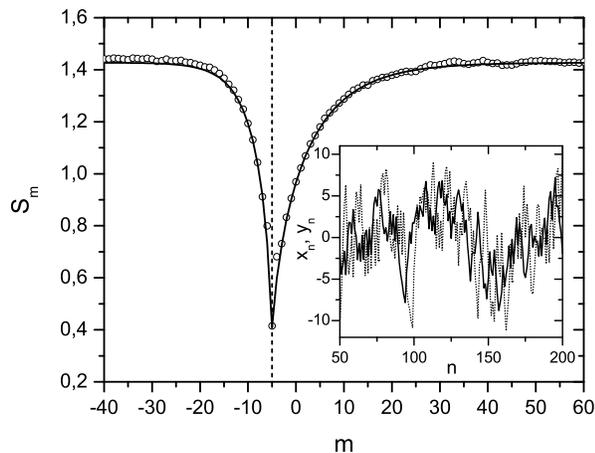}
\caption{Similarity function of the coupled delay maps Eq.~(\protect\ref%
{MismatchCoseno}) (circles). The parameters are $n_{0}=15,$ $a=0.9,$ $b=3$
and $n_{0}^{\prime }=20,$ $a^{\prime }=0.8,$ $b^{\prime }=5.$ The line
corresponds to the fitting Eq.~(\protect\ref{SimilMismatchIkeda}) joint with
the mapping Eq.~(\protect\ref{MappingMismacht}) [with $\protect\delta t=1]$,
indistinguishable from Eq.~(\protect\ref{SimilMapMismacht}). The inset
corresponds to a realization of $x_{n}$ (full line), and $y_{n}$ (dotted
line).}
\end{figure}

Clearly, the Langevin approach provides a very well fitting to the numerical
simulations. Furthermore, it allows to characterize the influence of the
parameters mismatch on the synchronization of the transmitter and receiver
variables. The analytical quantification of this effect can be obtained from
the similarity function Eq.~(\ref{SimilMismatchIkeda}) [or Eq.~(\ref%
{SimilMapMismacht})] evaluated at the origin
\end{subequations}
\begin{equation}
S(0)=\left[ \sqrt{\frac{\gamma ^{\prime }}{\alpha \gamma }}+\sqrt{\frac{%
\alpha \gamma }{\gamma ^{\prime }}}-\frac{4\sqrt{\gamma \gamma ^{\prime }}}{%
(\gamma +\gamma ^{\prime })}\exp [-\bar{\gamma}|\Delta |]\right] ^{1/2}.
\end{equation}%
By a direct inspection of this expression, we realize that the dependence of 
$S(0)$ on the prime parameters is non-monotonous, $S(0)$ develops different
minima when varying the receiver parameters. In agreement with the results
of Ref.~\cite{colet}, assuming that it is possible to adjust a given
receiver parameter, the previous expression allows us to choose the best
value that maximizes the synchronization phenomenon.

\section{Synchronization of second-order non-linear delay differential
equations}

In the previous sections, we analyzed the phenomenon of chaotic
synchronization for dynamics generated by first order delay equations.
Nevertheless, second order delay differential equations also arise in the in
the description of real experimental setups. By second order we mean
equations whose linear dynamical contributions are equivalent to second
order (time) derivative evolutions. Here, we demonstrate that the Langevin
approach also works in that case.

Following Ref.~\cite{colet} we consider the evolution%
\begin{eqnarray}
x(t)\!+\!\frac{\dot{x}(t)}{\gamma }\!+\!\frac{1}{\theta }\!\int_{0}^{t}%
\!x(s)ds\! &=&\!\beta \!\cos ^{2}[x(t-T)]\!+\!\xi _{x}(t),  \label{OscX} \\
y(t)\!+\!\frac{\dot{y}(t)}{\gamma ^{\prime }}\!+\!\frac{1}{\theta ^{\prime }}%
\!\int_{0}^{t}\!y(s)ds\! &=&\!\beta ^{\prime }\!\cos ^{2}[x(t-T^{\prime
})]\!+\!\xi _{y}(t),\ \ \ \ \   \label{OscY}
\end{eqnarray}%
which describe synchronization in a set of coupled \textit{electro-optical
laser devices}. The integral contributions proportional to $(1/\theta )$ and 
$(1/\theta ^{\prime })$ indicate that the linear evolutions defined by the
left hand side of Eqs.~(\ref{OscX}) and (\ref{OscY}) are equivalent to a set
of second order derivative differential equations. In addition to the
parameters mismatch, we also consider the action of external additive noises 
$\xi _{x}(t)$ and $\xi _{y}(t)$, whose mutual and self-correlations are
defined by Eq.~(\ref{ExternalNoiseCorrelation}). In order to simplify the
final expression here we not consider any mismatch in the phase of the
chaotic forces \cite{nota}.

Under the same conditions than in the previous sections, we assume that $x(t)
$ and $y(t),$ in the long time limit, attain the fully-developed chaos
regime, allowing us to replace the nonlinear delay forces by noise
contributions with the same time arguments, delivering 
\begin{subequations}
\label{OscilatoriaLangevin}
\begin{eqnarray}
x(t)\!+\!\frac{\dot{x}(t)}{\gamma }\!+\!\frac{1}{\theta }\int_{0}^{t}x(s)ds%
\! &=&\!\eta (t-T)\!+\!\xi _{x}(t), \\
y(t)\!+\!\frac{\dot{y}(t)}{\gamma ^{\prime }}\!+\!\frac{1}{\theta ^{\prime }}%
\int_{0}^{t}y(s)ds\! &=&\!\sqrt{\alpha }\eta (t-T^{\prime })\!+\!\xi
_{y}(t),\ \ \ \ \ \ \ 
\end{eqnarray}%
where as before, $\alpha =(\beta ^{\prime }/\beta )^{2}.$ The correlation of 
$\eta (t)$ is again defined by Eq.~(\ref{D_Correlation}), and we take $%
\overline{\eta (t)}=0.$ Notice that in spite that the nonlinear contribution 
$\cos ^{2}[x]$ is always positive, when $\{\theta ,\theta ^{\prime
}\}<\infty ,$ the second order linear evolution introduces self dynamical
oscillations that imply that its effective action averaged over realizations
(or its stationary time average) must be taken as zero. This property breaks
down when the integral contributions are absent, i.e., in the limit $\theta
=\theta ^{\prime }=\infty .$

The Green functions associated to the linear evolutions Eq.~(\ref%
{OscilatoriaLangevin}) can be obtained straightforwardly in the Laplace
domain, being defined by the addition of two exponential functions. After
integrating formally both equations for each realization of the noises, it
follow
\end{subequations}
\begin{subequations}
\label{Quadrado}
\begin{eqnarray}
\lim_{t\rightarrow \infty }\langle \overline{x^{2}(t)}\rangle  &=&\frac{%
\gamma }{2}(\mathcal{A}+\mathcal{A}_{xx}), \\
\lim_{t\rightarrow \infty }\langle \overline{y^{2}(t)}\rangle  &=&\frac{%
\gamma ^{\prime }}{2}(\alpha \mathcal{A}+\mathcal{A}_{yy}).
\end{eqnarray}
\end{subequations}
The similarity function [Eq.~(\ref{Similaridad})] reads 
\begin{widetext}
\begin{equation}
S(\tau )=\left\{ \frac{\sqrt{\frac{\gamma }{\alpha \gamma ^{\prime }}}\left(
1+\frac{\mathcal{A}_{xx}}{\mathcal{A}}\right) +\sqrt{\frac{\alpha \gamma
^{\prime }}{\gamma }}\left( 1+\frac{\mathcal{A}_{yy}}{\alpha \mathcal{A}} 
\right) -4\mu \left[ \Xi (\tau -\Delta )+\frac{\mathcal{A}_{xy}}{\sqrt{ 
\alpha }\mathcal{A}}\Xi (\tau )\right] }{\left( 1+\frac{\mathcal{A}_{xx}}{ 
\mathcal{A}}\right) ^{1/2}\left( 1+\frac{\mathcal{A}_{yy}}{\alpha \mathcal{A}} 
\right) ^{1/2}}\right\} ^{1/2}.  \label{SimilOscilatorio}
\end{equation}
\end{widetext}Here, $\Delta =T-T^{\prime }.$ The function $\Xi (t)$ is
defined as 
\begin{equation}
\Xi (t)=e^{-\frac{1}{2}\bar{\gamma}|t|}\left[ \cosh (\bar{\Phi}|t|/2)-\bar{%
\nu}\frac{\bar{\gamma}}{\bar{\Phi}}\sinh (\bar{\Phi}|t|/2)\right] ,
\end{equation}%
where the dissipative rate $\bar{\gamma}$ is defined by 
\begin{equation}
\bar{\gamma}\equiv \left\{ 
\begin{array}{c}
\gamma \ \ \ if\ \ \ t>0 \\ 
\gamma ^{\prime }\ \ \ if\ \ \ t<0%
\end{array}%
\right. .
\end{equation}%
For $t>0,$ the frequency $\bar{\Phi}$ reads 
\begin{equation}
\bar{\Phi}\equiv \gamma \sqrt{1-\frac{4}{\theta \gamma }},\ \ \ \ \ \ if\ \
\ t>0,  \label{Frequencia}
\end{equation}%
while the dimensionless parameter $\bar{\nu}$ is 
\begin{equation}
\bar{\nu}\equiv 1-\frac{2}{\theta (\theta +\theta ^{\prime })}\left( \frac{%
\theta }{\gamma }-\frac{\theta ^{\prime }}{\gamma ^{\prime }}\right) ,\ \ \
\ \ \ if\ \ \ t>0.
\end{equation}%
For $t<0,$ both $\Phi $ and $\nu $ are defined by interchanging $\theta
\leftrightarrow \theta ^{\prime }$ and $\gamma \leftrightarrow \gamma
^{\prime }$ in the previous expressions. Finally, in Eq.~(\ref%
{SimilOscilatorio}), we have also defined the dimensionless parameter%
\begin{equation}
\mu \equiv \frac{\sqrt{\gamma \gamma ^{\prime }}(\theta +\theta ^{\prime })}{%
(\theta \gamma ^{\prime }/\theta ^{\prime }\gamma )+[(\theta +\theta
^{\prime })(\gamma +\gamma ^{\prime })-2]+(\theta ^{\prime }\gamma /\gamma
^{\prime }\theta )},
\end{equation}%
which is symmetric in the emitter and receiver parameters. 
\begin{figure}[t]
\includegraphics[height=6cm,bb=14 15 298 228]{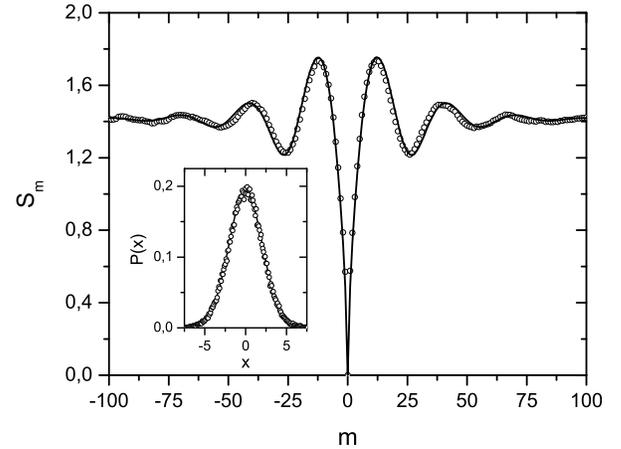}
\caption{Similarity function of the coupled delay maps Eq.~(\protect\ref%
{MapOsc}) (circles) without noises, $A_{xx}=A_{yy}=A_{xy}=0$ and without
mismatching, $n_{0}^{\prime }=n_{0},$ $a^{\prime }=a,$ $b^{\prime }=b$ and $%
\protect\omega ^{\prime }=\protect\omega .$ The parameters are $n_{0}=20,$ $%
a=0.9,$ $b=3$ and $\protect\omega =0.05.$ The line corresponds to the
fitting Eq.~(\protect\ref{SimilOscilatorio}) under the mapping Eq.~(\protect
\ref{MappingOscilador}) [with $\protect\delta t=1].$ The inset corresponds
to the stationary probability distribution of the process $x_{n}.$}
\label{FiguraChaos5}
\end{figure}

In order to check the validity of the previous approach, we consider the
discrete maps associated to Eqs.~(\ref{OscX}) and (\ref{OscY}) by Euler
integration 
\begin{subequations}
\label{MapOsc}
\begin{eqnarray}
x_{n+1} &=&ax_{n}-\omega \sum_{j=0}^{n}x_{j}+b\cos ^{2}(x_{n-n_{0}})+\xi
_{n}^{x}, \\
y_{n+1} &=&a^{\prime }y_{n}-\omega ^{\prime }\sum_{j=0}^{n}y_{j}+b^{\prime
}\cos ^{2}(x_{n-n_{0}^{\prime }})+\xi _{n}^{y}.\ \ \ \ \ 
\end{eqnarray}%
The (transmitter) parameters mapping reads
\end{subequations}
\begin{equation}
\gamma =\frac{(1-a)}{\delta t},\ \ \ \ \ \ \ \theta =\delta t\frac{(1-a)}{%
\omega },\ \ \ \ \ \ \ \beta =\frac{b}{(1-a)}.  \label{MappingOscilador}
\end{equation}%
The same relations are valid for the prime (receiver) parameters. The noises 
$\xi _{n}^{x}$ and $\xi _{n}^{y},$ joint with the corresponding mapping for
their amplitudes are defined below Eq.~(\ref{NoiseDrivenMap}). The chaotic
force amplitude of the continuous and discrete time evolutions are related
by $\mathcal{A}=A/\delta t.$

In Fig. 5 we plot the similarity function [Eq.~(\ref{SimilarDiscreta})]
associated to the coupled maps Eq.~(\ref{MapOsc}) in absence of parameters
mismatch and without the external noises. The analytical result Eq.~(\ref%
{SimilOscilatorio}), under the mapping Eq.~(\ref{MappingOscilador}),
provides an excellent fitting of the numerical results. In contrast with the
previous sections (see for example Fig. 1), here the similarity function
develops an oscillatory behavior, its origin can be associated to the
integral contributions in the dissipative dynamics of Eqs.~(\ref{OscX}) and (%
\ref{OscY}). Their characteristic frequency follows from Eq.~(\ref%
{Frequencia}).

In order to check the achievement of the fully-developed chaos regime \cite%
{kilom}, in the inset we show the stationary probability distribution of $%
x_{n}.$ Consistently, this distribution can be fitted with a Gaussian
distribution.

In Fig. 6, for other set of characteristic parameters, we plot the
similarity function in presence of parameter mismatch and external noise
sources. As in the previous case, the fitting Eq.~(\ref{SimilOscilatorio})
correctly match the numerical results. The effective chaotic force amplitude
is $A\approx 1.2.$ The asymmetry of $S_{m}$ around its minimum value follows
from the parameter mismatch between the evolution of $x_{n}$ and $y_{n}.$

In the inset, maintaining the parameters corresponding to the evolution of $%
x_{n},$ we plot the similarity function in absence of both, parameters
mismatch and the external noise contributions. In this case, the chaotic
force amplitude [determine from Eq.~(\ref{Quadrado})] is $A\approx 1.$ 
\begin{figure}[t]
\includegraphics[height=6cm,bb=14 15 298 228]{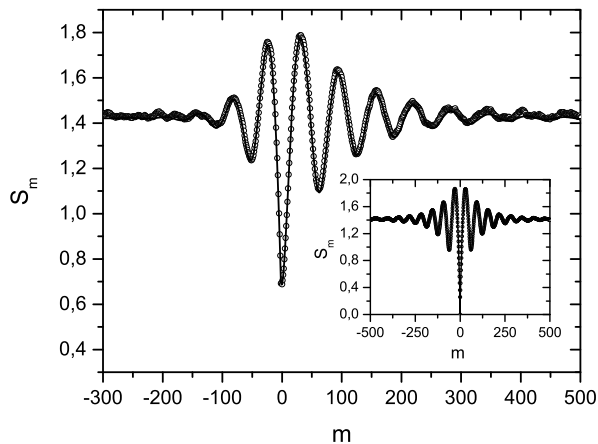}
\caption{Similarity function of the coupled delay maps Eq.~(\protect\ref%
{MapOsc}) (circles) with parameters mismatch and under the action of
external noises. The parameters are $n_{0}=20,$ $a=0.98,$ $b=3,$ $\protect%
\omega =0.01,$ and $n_{0}^{\prime }=22,$ $a^{\prime }=0.95,$ $b^{\prime }=4,$
$\protect\omega ^{\prime }=0.0125.$ The noise parameters are $%
A_{xx}=A_{yy}=0.0225$ and $A_{xy}=0.$ The line corresponds to the fitting
Eq.~(\protect\ref{SimilOscilatorio}) joint with the mapping Eq.~(\protect\ref%
{MappingOscilador}) [with $\protect\delta t=1].$ The inset corresponds to
the similarity function in absence of noises and without mismatching.}
\label{FiguraChaos6}
\end{figure}

As in the previous section, a remarkable aspect of our analytical results is
that they allows us to know the influence of different realistic effects on
the chaotic synchronized manifold. For example, for a given set of fixed
parameters, which may include the amplitude of the external noises, one can
determine the value of the rest of the parameters that minimizes the
similarity function at $\tau =0,$ giving rise to a maximization in the
degree of synchronization between the emitter and receiver variables. This
kind of analysis follows straightforwardly from our analytical results.

\section{Conclusions}

In this paper, we have characterized the phenomenon of chaotic
synchronization in scalar coupled non-linear time delay dynamics. Our
formalism relies in recognizing that in a fully-developed chaos regime, the
trajectories associated to a broad class of hyperchaotic attractors are
statistically equivalent to the realizations of a linear Langevin equation.
This equivalence can be established when the function that defines the
driven chaotic delay force is an oscillatory one, such that its dynamical
action can be represented by a delta-correlated noise. Given this Langevin
representation, the coupled nonlinear delay evolutions, where the phenomenon
of chaotic synchronization happens, are replaced by a set of linear
stochastic equations where the noise that represents the chaotic force
maintains the corresponding time arguments. The statistical properties of
the Langevin equations can be obtained analytically, providing an excellent
estimation of the stationary statistical properties of the synchronized
manifold.

Using the Langevin representation, we analyzed the phenomenon of anticipated
synchronization in the presence of external additive noises. We also
characterized the effect of parameters mismatch in a hyperchaotic
communication scheme. Second-order delay equations associated to an electro
optical device were also characterized. The analytical predictions of the
Langevin approach correctly fit numerical simulations in discrete coupled
nonlinear delay maps associated to the corresponding continuous time
evolutions.

In all cases, the degree of synchronization (between the master-slave or
emitter-receiver variables) was characterized through a similarity function,
defined in terms of the correlation between the synchronizing systems. When
the departure from perfect synchronization is due to a parameter
mismatching, the fitting to the similarity function does not involve any
free parameter. When the action of external noises is considered the fitting
depends on the effective chaotic force amplitude.

Our results are interesting from both, theoretical and experimental point of
view. From our analytical expressions it is possible to evaluate under which
conditions a given undesired effect can be minimized by controlling the rest
of the parameters. On the other hand, in the context of hyperchaotic
communication schemes, while high dimensional systems increase the
complexity of the masking signals, our results show that the corresponding
statistical properties may adopt a simple analytical form. In fact, by
measuring the similarity function our results allow to infer the value of
some of the characteristic parameters of the hyperchaotic delay dynamics.

The present study may be continued in different relevant directions such as
the extension of the Langevin representation beyond the fully-developed
chaos regime (non-Gaussian chaos) as well as for non-scalar chaotic
dynamics. Furthermore, the characterization of the dependence of the
effective chaotic force amplitude with the external noise parameters is an
open interesting issue.

\section*{Acknowledgments}

The author thanks fruitful discussions with Prof. D.H. Zanette at Centro At%
\'{o}mico Bariloche, as well as with Prof. G. Buendia at Consortium of the
Americas for Interdisciplinary Science. This work was supported by CONICET,
Argentina.


\begin{thebibliography}{99}
\bibitem{notaReferee} There exist cases where this condition is not
satisfied, as for example for the Mackey-Glass dynamic \cite{mackey,berre1},
which arise in the context of white-blood-cell production.
\bibitem{nota} It is also possible to consider extra parameters mismatch in
the transmitter-receiver nonlinearities, as for example the phase $\phi $ in 
$f(x)=\sin (x+\phi )$ \cite{colet}. This case can be worked out in the
Langevin representation in a similar way by discomposing the $\sin $
function in two shiftless terms and considering their respective symmetries
properties. For simplicity, we will not consider this case.
\bibitem{kilom} Taking the experimental parameters of Ref. \cite{colet}, the
dynamics goes beyond the Gaussian chaos regime. Our formalism may also be
used as an estimator of the statistical properties in this regime.
Nevertheless, the quality of the fitting departs from that showed in the
figures of this paper. On the other hand, from Eq.~(\ref{OscilatoriaLangevin}%
), it is possible to write the Fourier transform ($t\rightarrow w$) of $x(t)$
as $x(w)=\{i\gamma w/[-w^{2}+i\gamma w+\gamma /\theta )]\}[\eta (w)+\xi
_{x}(w)],$ and a similar one for $y(w).$ Then, the band-pass filter
approximation introduced in Ref. \cite{colet} can be read as a rough
approximation to these expressions.
\end{thebibliography}
\end{document}